# Efficient Power System Transient Simulation Based on Frequency Response Optimized Integrators Considering Second Order Derivative

Sheng Lei, Student Member, IEEE, and Alexander Flueck, Senior Member, IEEE
Department of Electrical and Computer Engineering
Illinois Institute of Technology
Chicago, IL, USA
Email: slei3@hawk.iit.edu, and flueck@iit.edu

*Abstract*—Frequency response optimized integrators considering second order derivative are proposed in this paper. Based on the proposed numerical integrators, and others which also consider second order derivative, this paper puts forward a novel power system transient simulation scheme. Instead of using a unique numerical integrator, the proposed simulation scheme chooses proper ones according to the dominant frequency component of the differential state variables. With the proposed simulation scheme, computational efficiency is improved by using large step sizes without sacrificing accuracy. Numerical case studies demonstrate the validity and efficiency of the simulation scheme.

## I. INTRODUCTION

Modern power systems have witnessed significant increase in the amount of distributed energy resource (DER) integration due to environmental and economic considerations. With the already high penetration, the devices through which DERs are integrated considerably impact overall power system dynamics and stability. Many of these devices are single-phase ones. Moreover, they are usually installed in structurally unbalanced distribution systems. As a result, more and more research efforts are devoted to stability study on unbalanced power systems [1]-[3]. Unfortunately such a problem is out of the capability of positive-sequence transient stability (TS) simulation traditionally used in stability study on transmission systems, which assumes three-phase balance [4].

Electromagnetic transient (EMT) simulation models individual phases in detail. It is a natural alternative for stability study on unbalanced power systems. This type of study has already been reported in the literature [2], [3]. However small step sizes of tens to hundreds of microseconds are typically used with EMT simulation, making it computationally inefficient and time-consuming [4], [5].

For power system stability study, the electromechanical dynamics, after fast transients have died down, are of interest to determine whether the system regains stable operation [6], [7]. Lumped parameter models of power system networks and devices are adopted [6], [7]. Consequently millisecond-level step sizes should be applicable for studying such dynamics. In some existing EMT simulators, network equations and machine dynamics equations are solved separately by introducing an artificial delay, which injects a step size dependent error into the solution process, leading to distortion of results or even divergence [8], [9]. This constraint on step size can be relaxed by solving the whole system simultaneously via an iterative method. Another factor which constrains step size is the numerical error caused by the implicit trapezoidal method broadly adopted by existing EMT simulators with large step sizes [10], [11], which is to be dealt with in this paper.

In order to reduce numerical error and enable large step sizes for efficient transient simulation, this paper proposes a simulation scheme based on frequency response optimized integrators considering second order derivative. Contributions of this paper are two-fold. First, it proposes frequency response optimized integrators considering second order derivative. These numerical integrators effectively reduce the numerical error introduced by discretizing ordinary differential equations. Other numerical integrators considering second order derivative can also be reviewed from the new perspective of frequency response. Second, the implementation of the proposed simulation scheme is introduced in this paper. In the proposed simulation scheme, numerical integrators are chosen according to the spectral properties of the state variable of individual ordinary differential equations. Network equations are solved together with equations of other power system devices so that consistency is guaranteed. Discontinuities such as switching events are properly dealt with.

The rest of the paper is organized as follow. Section II defines the frequency response optimized integrators considering second order derivative. It also reviews other numerical integrators considering second order derivative from the frequency response perspective. Section III introduces the implementation of the proposed simulation scheme. Numerical case studies verifying the validity and efficiency of the proposed simulation scheme appear in Section IV. Section V concludes the paper and points out some directions for future research.

## II. Frequency Response Optimized Integrators Considering Second Order Derivative

In order to achieve high efficiency in power system transient simulation, numerical integrators should be frequency response optimized according to the spectral properties of power system signals, so that the constraint on the step size selection can be relaxed from the numerical error point of view. Some power system signals, for example AC voltages and currents, are typically around nominal fundamental angular frequency $\omega_{sync}$. The numerical integrators for differential equations of these signals should be designed so that they introduce only slight numerical error to signals around $\omega_{sync}$ despite the step size. On the other hand, some power system signals are slow variations, examples of which include rotor angle relative to synchronously rotating reference frame (referred to as rotor angle hereafter for simplicity) and rotor speed of synchronous generators. The numerical integrators for differential equations of these signals should be highly accurate for signals around 0 Hz.

Consider a general ordinary differential equation of the form

$$\dot{x} = f(x,u) \tag{1}$$

where $x$ denotes the state variable; $u$ denotes the input; $f$ is a function depending on $x$ and $u$. Although only a single input is considered here for simplicity, the same idea applies to multiple inputs and the further derivation is similar.

The traditional discretized version of (1) is [10], [11]

$$\begin{aligned} x_t &= x_{t-h} + \rho f(x_t,u_t) + \rho_c f(x_{t-h},u_{t-h}) \\ &= x_{t-h} + \rho \dot{x}_t + \rho_c \dot{x}_{t-h} \end{aligned} \tag{2}$$

where $x_t$ denotes the state variable value at the current time step; $x_{t-h}$ denotes the state variable value at the previous time step; $u_t$ denotes the input value at the current time step; $u_{t-h}$ denotes the input value at the previous time step; $h$ denotes the step size; $\rho$ and $\rho_c$ are coefficients to be determined. A specific selection for $\rho$ and $\rho_c$ determines a numerical integrator. For example, the implicit trapezoidal method [12] has $\rho = h/2$ and $\rho_c = h/2$; the backward Euler method [12] has $\rho = h$ and $\rho_c = 0$.

To achieve higher accuracy in computing $x_t$, second order derivative of $x$ may be considered

$$x_t = x_{t-h} + b_0 \dot{x}_t + b_{-1} \dot{x}_{t-h} + c_0 \ddot{x}_t + c_{-1} \ddot{x}_{t-h} \tag{3}$$

where $b_0$, $b_{-1}$, $c_0$ and $c_{-1}$ are coefficients to be determined. A specific selection for these coefficients determines a numerical integrator. Second order derivative of $x$ can be calculated by taking derivative on both sides of (1)

$$\ddot{x} = \frac{\partial f}{\partial x}\dot{x} + \frac{\partial f}{\partial u}\dot{u} = \frac{\partial f}{\partial x}f(x,u) + \frac{\partial f}{\partial u}\dot{u} \tag{4}$$

Substituting (4) into (3)

$$\begin{aligned} x_t &= x_{t-h} + b_0 f(x_t,u_t) + b_{-1} f(x_{t-h},u_{t-h}) \\ &+ c_0 (\frac{\partial f}{\partial x}|_t f(x_t,u_t) + \frac{\partial f}{\partial u}|_t \dot{u}_t) \\ &+ c_{-1}(\frac{\partial f}{\partial x}|_{t-h} f(x_{t-h},u_{t-h}) + \frac{\partial f}{\partial u}|_{t-h} \dot{u}_{t-h}) \end{aligned} \tag{5}$$

Obviously (5) or equivalently (3) requires derivative of $u$. Such information is known, or may be calculated analytically or numerically.

Values for the coefficients in (2) may be selected according to frequency response criteria [10], [11]. The similar idea is adopted in this paper to determine the coefficients in (3). Specifically performing Laplace transform on both sides of (3)

$$X = Xe^{-sh} + b_0 sX + b_{-1}sXe^{-sh} + c_0 s^2 X + c_{-1}s^2 X e^{-sh} \tag{6}$$

The $s$-domain error expression is

$$X - Xe^{-sh} - b_0 sX - b_{-1}sXe^{-sh} - c_0 s^2 X - c_{-1}s^2 X e^{-sh} \tag{7}$$

The $s$-domain relative error expression is

$$1 - e^{-sh} - b_0 s - b_{-1}se^{-sh} - c_0 s^2 - c_{-1}s^2 e^{-sh} \tag{8}$$

If the coefficients are chosen so that $s = j\omega_{select}$ and $s = -j\omega_{select}$ are both a root of the relative error expression (8), then the numerical integrator is accurate for signals at angular frequency $\omega_{select}$ despite the step size. Desirable selection is achieved by solving equations regarding the root conditions.

### A. Making $j\omega_{select}$ and $-j\omega_{select}$ a Single Root and 0 a Triple Root

This paper proposes a numerical integrator expressed as (9), where $j\omega_{select}$ and $-j\omega_{select}$ are a single root while 0 is a triple root of the relative error expression (8). This integrator is referred to as Integrator A hereafter.

$$\begin{aligned} b_0 &= \frac{h}{2}, \ b_{-1} = \frac{h}{2}, \ c_0 = -\frac{1}{\omega_{select}^2} + \frac{h}{2\omega_{select}}\cot(\frac{\omega_{select}h}{2}), \\ c_{-1} &= \frac{1}{\omega_{select}^2} - \frac{h}{2\omega_{select}}\cot(\frac{\omega_{select}h}{2}) \end{aligned} \tag{9}$$

By letting $\omega_{select} = \omega_{sync}$, Integrator A is accurate for signals at angular frequency $\omega_{sync}$ despite the step size. The integrator is also rather accurate for signals around 0 Hz.

### B. Making $j\omega_{select}$, $-j\omega_{select}$ and 0 a Single Root Respectively

This paper proposes a numerical integrator expressed as (10), where $j\omega_{select}$, $-j\omega_{select}$ and 0 are a single root of the relative error expression (8) respectively. This integrator is referred to as Integrator B hereafter.

$$b_0 = \frac{\sin(\omega_{select}h)}{\omega_{select}}, \; b_{-1} = 0, \; c_0 = \frac{\cos(\omega_{select}h)-1}{\omega_{select}^2}, \; c_{-1} = 0 \quad (10)$$

By letting $\omega_{select} = \omega_{sync}$, Integrator B is accurate for signals at angular frequency $\omega_{sync}$ and 0 Hz despite the step size. As $b_{-1} = 0$ and $c_{-1} = 0$, this integrator is suitable for dealing with discontinuities.

### C. Other Numerical Integrators Considering Second Order Derivative

A numerical integrator invented by Obrechkoff [13] is expressed as (11). From the frequency response perspective, 0 is a quintuple root of the relative error expression (8). Therefore it is highly accurate for signals around 0 Hz. This integrator is referred to as Integrator C hereafter.

$$b_0 = \frac{h}{2}, \; b_{-1} = \frac{h}{2}, \; c_0 = -\frac{h^2}{12}, \; c_{-1} = \frac{h^2}{12} \quad (11)$$

An implicit second order Taylor series method [12] is expressed as (12). From the frequency response perspective, 0 is a triple root of the relative error expression (8). Therefore it is rather accurate for signals around 0 Hz. As $b_{-1} = 0$ and $c_{-1} = 0$, it is suitable for dealing with discontinuities. The integrator is referred to as Integrator D hereafter.

$$b_0 = h, \; b_{-1} = 0, \; c_0 = -\frac{h^2}{2}, \; c_{-1} = 0 \quad (12)$$

The two numerical integrators reviewed in this subsection can also be understood as frequency response optimized integrators considering second order derivative in the sense that they are optimized for signals around 0 Hz and they do consider second order derivative of the state variable.

## III. PROPOSED SIMULATION SCHEME

### A. Power System Modeling

Transient models of power system equipment are well known and available in the literature [6]-[9]. These models consist of ordinary differential equations and algebraic equations, introducing differential state variables and algebraic state variables. The ordinary differential equations have to be discretized to be solved numerically. Instead of discretizing all of the ordinary differential equations with a unique numerical integrator, such as the implicit trapezoidal method used in traditional EMT simulation, the proposed simulation scheme discretizes them with multiple frequency response optimized integrators considering second order derivative, according to the spectral properties of the corresponding state variables.

As mentioned in Section II, some state variables are dominated by the fundamental frequency component. The ordinary differential equations of such state variables are discretized with Integrator A. On the other hand, some state variables are basically slow variants around 0 Hz. The ordinary differential equations of these state variables are discretized with Integrator C.

### B. Execution of the Proposed Simulation Scheme

The proposed simulation scheme adopts a fixed step size during a simulation run. This setting is the same as commonly used in EMT and TS simulators. The equations of power system network equipment are solved simultaneously with the equations of other power system devices via iteration at each time step. This simultaneous solution ensures consistency of the whole system to achieve high fidelity.

Discontinuities including faults and breaker switching should be dealt with carefully to obtain accuracy and to avoid possible numerical oscillation [14]. The proposed simulation scheme adopts an idea similar to Critical Damping Adjustment (CDA) [14] for these events. Immediately after a discontinuity, two half time steps are calculated with ordinary differential equations discretized with numerical integrators which have zero $b_{-1}$ and $c_{-1}$ as mentioned in Section II. For those ordinary differential equations discretized with Integrator A at normal time steps, Integrator B is used for discontinuities. For those discretized with Integrator C at normal time steps, Integrator D is used for discontinuities.

## IV. NUMERICAL CASE STUDIES

The proposed scheme is implemented with MATLAB. Its validity and efficiency is to be verified via numerical case studies in this section. To make fair comparison on accuracy and efficiency, an iterative EMT simulator is also implemented with MATLAB, adopting the same structure and execution as the proposed scheme, except that ordinary differential equations are discretized with the implicit trapezoidal method. Correctness of this iterative EMT simulator is validated in this section by comparing its results with the ones from SIMULINK. Note that this iterative EMT simulator solves the whole system simultaneously by iteration to maintain consistency so step sizes much larger than the typical 50 μs are possible.

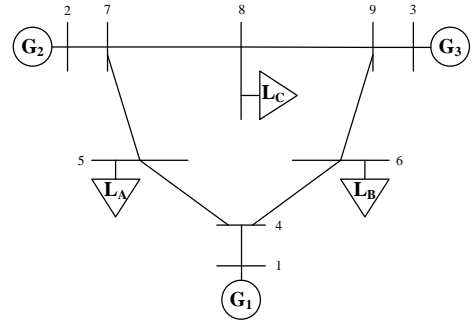

Figure 1.  3-machine 9-bus power system.

Numerical case studies are performed on a 3-machine 9-bus power system as shown in Fig. 1. Bus parameters, branch parameters and power flow data can be found in [15]. They are not given in this paper due to space limitation. Generator dynamic parameters used in this paper are given in the appendix. Unbalance is introduced by non-uniform allocation of loads on individual phases. In this paper, the total load at a specific bus is allocated as follows

$$S_A = \frac{1}{3}(1-k)S, \; S_B = \frac{1}{3}S, \; S_C = \frac{1}{3}(1+k)S \quad (13)$$

where $S$ is the total apparent power load at the bus; $S_A$, $S_B$ and $S_C$ are the Phase A, B and C apparent power load respectively; $k$ is an allocation factor, in this paper $k = 0.1$. Loads are assumed to be constant impedance in transient simulation.

At 0.1 s, a Phase-B-and-Phase-C-to-ground fault is applied at Bus 9 with a fault resistance of 0.001 p.u.. At 0.3 s, the fault is cleared. The same simulation is carried out with SIMULINK, the iterative EMT simulator and the proposed scheme.

### A. Validation of the Iterative EMT Simulator

The SIMULINK benchmark system uses the variable-step solver ode23t with a relative tolerance of $10^{-3}$ and a max step size of 50 μs. The iterative EMT simulator uses a step size of 50 μs. Results from the iterative EMT simulator are compared with the ones from SIMULINK. Bus 9 Phase A voltage and Generator 3 rotor speed are shown in Figs. 2 and 3 respectively. Note that in each figure, the two curves basically overlap each other. This high degree agreement of the two simulators verifies the correctness of the iterative EMT simulator.

### B. Validity and Efficiency of the Proposed Scheme

Results from the iterative EMT simulator with a tiny step size of 5 μs are used as the reference. Results from the proposed scheme and the iterative EMT simulator with larger step sizes are compared to the reference so that the numerical error of these two schemes can be observed. Figs. 4 and 5 show Generator 3 rotor angle. Note that the proposed scheme gives more accurate results than the iterative EMT simulator with the same or one-half the step size. It is thus possible to use the proposed scheme with large step sizes to reduce the time consumption, while maintaining satisfactory accuracy.

To quantitatively study the numerical error, this paper uses the following error measurement. Suppose $x$ is a variable to be considered, the relative error regarding $x$ of a simulation scheme with a specified step size is defined as

$$err(x) = \|x_{com} - x_{ref}\|_2 / \|x_{ref}\|_2 \times 100 \quad (14)$$

where $x_{com}$ is the computed value from a simulation scheme with a specified step size; $x_{ref}$ is the reference value. As results from digital simulation are discrete, the 2-norm is calculated at common time instants of $x_{com}$ and $x_{ref}$.

This paper considers averaged relative error of nodal voltages and rotor angles. These two errors are referred to as voltage error and rotor angle error respectively hereafter for simplicity. Specifically, voltage error is defined as

$$ERR(v) = \frac{1}{n_{node}} \sum_{i=1}^{n_{node}} err(v_i) \quad (15)$$

where $v_i$ is the voltage at Node $i$; $n_{node}$ is the total number of nodes. Note that in multi-phase power systems, a bus may have multiple nodes. Similarly rotor angle error is defined as

$$ERR(\delta) = \frac{1}{n_{gen}} \sum_{j=1}^{n_{gen}} err(\delta_j) \quad (16)$$

where $\delta_j$ is the rotor angle of Generator j; $n_{gen}$ is the total number of generators. Voltage error and rotor angle error of the proposed scheme and the iterative EMT simulator with different step sizes, as well as their time consumption, are listed in Table I. Duration of the simulations to generate Table I is from 0.0 s to 2.0 s.

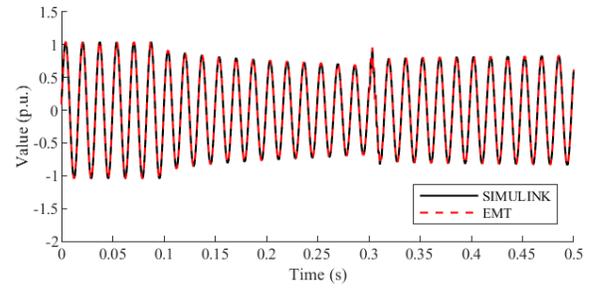

Figure 2. Bus 9 Phase A voltage. Black solid line: SIMULINK. Red dashed line: the iterative EMT simulator.

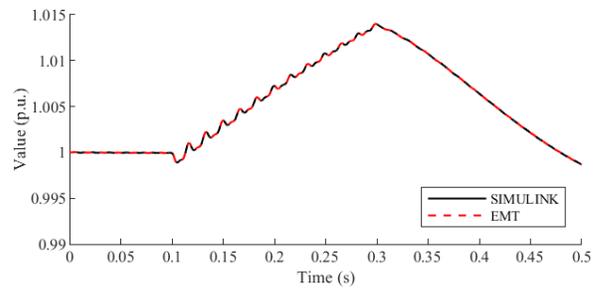

Figure 3. Generator 3 rotor speed. Black solid line: SIMULINK. Red dashed line: the iterative EMT simulator.

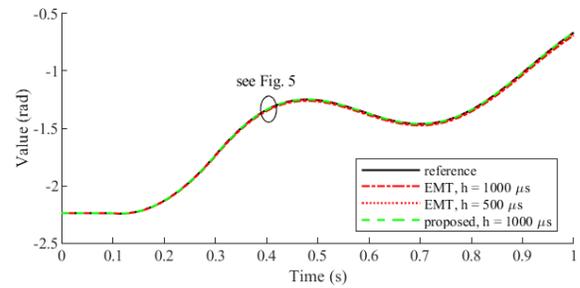

Figure 4. Generator 3 rotor angle. Black solid line: the reference. Red dash-dot line: the iterative EMT simulator with a 1000 μs step size. Red dotted line: the iterative EMT simulator with a 500 μs step size. Green dashed line: the proposed scheme with a 1000 μs step size.

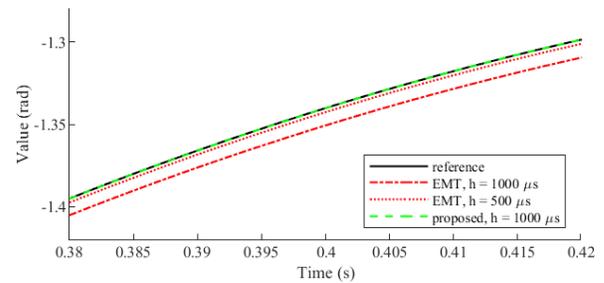

Figure 5. Generator 3 rotor angle (zoomed in). Black solid line: the reference. Red dash-dot line: the iterative EMT simulator with a 1000 μs step size. Red dotted line: the iterative EMT simulator with a 500 μs step size. Green dashed line: the proposed scheme with a 1000 μs step size.

TABLE I. COMPARISON OF THE PROPOSED SCHEME AND THE ITERATIVE EMT SIMULATOR

| Step Size (µs) | Time Consumption (s) | | Voltage Error | | Rotor Angle Error | |
|---|---|---|---|---|---|---|
| | Proposed | EMT | Proposed | EMT | Proposed | EMT |
| 125 | 148.52 | 65.78 | 0.0071 | 0.3393 | 0.0000 | 0.0414 |
| 250 | 74.74 | 33.57 | 0.0377 | 1.0454 | 0.0001 | 0.1655 |
| 500 | 39.85 | 16.63 | 0.2816 | 2.1341 | 0.0012 | 0.6676 |
| 1000 | 24.41 | 8.66 | 1.0487 | 5.0507 | 0.0076 | 2.7692 |
| 2000 | 12.20 | 4.42 | 1.2813 | 20.564 | 0.1436 | 12.625 |
| 4000 | 6.31 | 2.55 | 2.2874 | 119.135 | 1.3135 | 84.877 |

As can be seen from Table I, the proposed scheme requires a little more than twice the time consumption of the iterative EMT simulator with the same step size. This higher computational cost is due to the additional consideration of second order derivative by the numerical integrators. Nevertheless the increased cost pays off with much higher accuracy which can be 4 times compared to the iterative EMT simulator. Therefore computational efficiency is achieved by the proposed scheme using larger step sizes to obtain similar or even higher accuracy with less time consumption. For example, the iterative EMT simulator takes 16.63 s to complete the simulation with a 500 µs step size. However the proposed scheme takes only 12.20 s with a 2000 µs step size, and still gives more accurate results.

As a special EMT simulation scheme, the proposed one can be applied to any problems that this type of simulation is fit for. However the proposed scheme is especially suitable for stability study on unbalanced power systems. As is shown in Table I, the proposed scheme provides sufficient accuracy with millisecond-level step sizes, while the error of the iterative EMT simulator is considerable. The selection of step sizes is problem-dependent. Based on the many computational experiments that the authors have performed, a step size of 1 or 2 ms (1000 or 2000 µs) is a good trade-off between efficiency and accuracy for stability study on unbalanced power systems.

## V. CONCLUSION AND FUTURE WORK

A power system transient simulation scheme based on frequency response optimized integrators considering second order derivative is put forward to achieve computational efficiency. Frequency response optimized integrators considering second order derivative are defined in this paper. Other numerical integrators are reviewed from the frequency response perspective. The implementation of the proposed scheme is introduced. The validity and efficiency of the proposed scheme is verified by numerical case studies.

In the future, more power system equipment models may be added to extend the functionality of the proposed scheme. Parallel implementation may be investigated to further increase its efficiency. Research efforts may also be directed to applications of the proposed scheme, such as combined transmission-distribution simulation and sub-synchronous resonance study.

## APPENDIX

TABLE A. I. GENERATOR DYNAMIC PARAMETERS

| Generator (#) | 1 | 2 | 3 |
|---|---|---|---|
| Rated Power (MVA) | 247.5 | 192.0 | 128.0 |
| Stator Resistance (p.u.) | 0.002 | 0.002 | 0.002 |
| Stator Leakage Reactance (p.u.) | 0.0787 | 0.0787 | 0.0787 |
| D-Axis Synchronous Reactance (p.u.) | 1.575 | 1.575 | 1.575 |
| Q-Axis Synchronous Reactance (p.u.) | 1.512 | 1.512 | 1.512 |
| D-Axis Transient Reactance (p.u.) | 0.291 | 0.291 | 0.291 |
| Q-Axis Transient Reactance (p.u.) | 0.39 | 0.39 | 0.39 |
| D-Axis Subtransient Reactance (p.u.) | 0.1733 | 0.1733 | 0.1733 |
| Q-Axis Subtransient Reactance (p.u.) | 0.1733 | 0.1733 | 0.1733 |
| Open-Circuit D-Axis Transient Time Constant (s) | 6.1 | 6.1 | 6.1 |
| Open-Circuit Q-Axis Transient Time Constant (s) | 1.0 | 1.0 | 1.0 |
| Open-Circuit D-Axis Subtransient Time Constant (s) | 0.05 | 0.05 | 0.05 |
| Open-Circuit Q-Axis Subtransient Time Constant (s) | 0.15 | 0.15 | 0.15 |
| Inertia Constant (s) | 23.64 | 6.4 | 3.01 |
| Damping Coefficient (p.u.) | 0.1 | 0.1 | 0.1 |